\documentclass[conference]{IEEEtran}
\IEEEoverridecommandlockouts

\usepackage{cite}
\usepackage{amsmath,amssymb,amsfonts}
\usepackage{algorithmic}
\usepackage{graphicx}
\usepackage{textcomp}
\usepackage{xcolor}
\usepackage{multirow}
\usepackage{hyperref}
\usepackage[top=0.75in, bottom=1in, left=0.625in, right=0.625in]{geometry}

\def\BibTeX{{\rm B\kern-.05em{\sc i\kern-.025em b}\kern-.08em
    T\kern-.1667em\lower.7ex\hbox{E}\kern-.125emX}}
\begin{document}

\title{DACN: Dual-Attention Convolutional Network for Hyperspectral Image Super-Resolution\\
}

\author{
    \IEEEauthorblockN{
        Usman Muhammad\textsuperscript{1} and Jorma Laaksonen\textsuperscript{1}
    }
    \IEEEauthorblockA{\textsuperscript{1} Department of Computer Science, Aalto University, Finland}
}


\maketitle

\begin{abstract}
2D convolutional neural networks (CNNs) have attracted significant attention for hyperspectral image super-resolution tasks. However, a key limitation is their reliance on local neighborhoods, which leads to a lack of global contextual understanding. Moreover, band correlation and data scarcity continue to limit their performance. To mitigate these issues, we introduce DACN, a dual-attention convolutional network for hyperspectral image super-resolution. Specifically, the model first employs augmented convolutions, integrating multi-head attention to effectively capture both local and global feature dependencies. Next, we infer separate attention maps for the channel and spatial dimensions to determine where to focus across different channels and spatial positions. Furthermore, a custom optimized loss function is proposed that combines L2 regularization with spatial-spectral gradient loss to ensure accurate spectral fidelity. Experimental results on two hyperspectral datasets demonstrate that the combination of multi-head attention and channel attention outperforms either attention mechanism used individually. The source codes are publicly available at: \href{https://github.com/Usman1021/dual-attention}{https://github.com/Usman1021/dual-attention}.
\end{abstract}

\begin{figure*}[htbp]
    \centering
    \includegraphics[width=\textwidth]{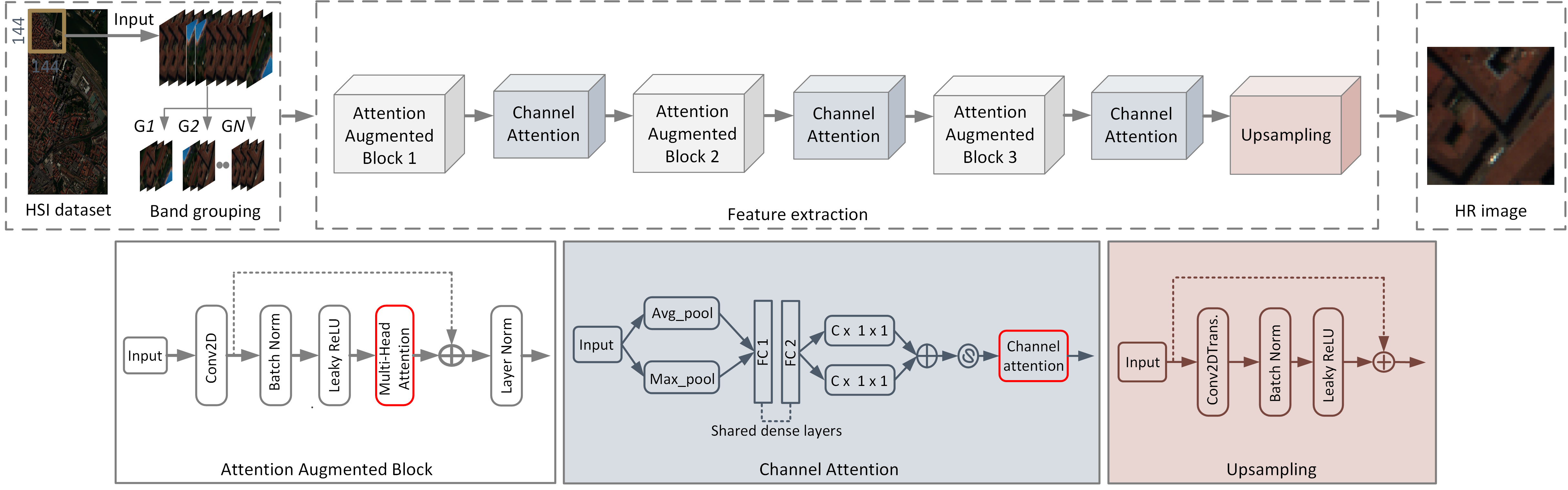} 
    \caption{An overview of the proposed DACN model: the white block on the left illustrates the integration of multi-head attention, while the gray block in the middle emphasizes channel attention. The beige-colored block on the right represents the upsampling module, which uses transposed convolution with a skip connection.}
    \label{fig}
\end{figure*}

\section{Introduction}
Hyperspectral images (HSIs) typically offer high spectral resolution but suffer from low spatial resolution due to hardware constraints, while multispectral images (MSIs) generally have lower spectral resolution but higher spatial resolution \cite{li2019hyperspectral}. The wide spectrum of hyperspectral images is extremely valuable for a variety of applications, including Earth observation, forest monitoring, and satellite image scene classification \cite{muhammad2018pre, muhammad2018feature, muhammad2019bag, muhammad2022patch}. The primary goal of the single image super-resolution (SR) task is to enhance a degraded low-resolution (LR) image by reconstructing its corresponding high-resolution (HR) version. In the early years, single image super-resolution methods were mainly based on interpolation techniques, such as nearest neighbor, bilinear, and bicubic interpolation \cite{jiang2020single}. These methods were computationally simple and well suited to the hardware capabilities of that time. However, since they relied solely on computations of pixel value without considering image content or prior information, reconstructed high-resolution images often lacked fine details \cite{li2019hyperspectral}. 

In recent years, convolutional neural network (CNN)-based methods have emerged as the dominant approach for super-resolution tasks \cite{hou2022deep, zhang2024hyperspectral, li2020mixed, zhang2021multi}. CNN-based methods extract image features using shared weighted convolutional kernels, which exhibit local connectivity and translation invariance. Although these characteristics improve the efficiency and generalization of CNNs, they also introduce two key limitations: (a) convolution kernels are restricted by their local receptive fields, making it difficult to capture long-range pixel dependencies in images; and (b) the static weights of convolution kernels during inference prevent them from dynamically adapting to the input content \cite{zhao2024ssir}.

In contrast, self-attention has recently gained recognition as an effective mechanism for capturing long-range dependencies in data. It has been particularly impactful in sequence modeling and generative tasks, such as natural language processing and machine translation \cite{vaswani2017attention}. The core principle of self-attention involves computing a weighted sum of input representations, where the weights are dynamically assigned based on the similarity between different positions in the input sequence \cite{bello2019attention}. This flexible weighting helps the model focus on important parts of the input while processing, unlike pooling or convolution, which use fixed weights and limited areas. In addition, attention \cite{woo2018cbam} modules have been extensively studied in previous literature, not only determining where to focus, but also enhancing the representation of important features \cite{mnih2014recurrent, ba2014multiple}. 

Although attention-based models demonstrate excellent performance on multiple benchmarks (ImageNet-1K, MS COCO, and VOC 2007) \cite{woo2018cbam, hu2018squeeze}, most previous work uses attention for task-specific purposes \cite{park2018bam}. 
In contrast, the super-resolution task requires low-resolution images that are typically not downsampled during feature extraction, as this would lead to further information loss, making HR reconstruction more challenging \cite{zhao2024ssir}. We argue that the potential of self-attention and attention mechanisms needs further exploration for hyperspectral SR tasks. Therefore, to effectively capture long-range pixel dependencies and model global relationships, a key question needs to be answered: How would performance be impacted by incorporating self-attention with an attention mechanism into convolutional networks for hyperspectral super-resolution tasks?

Motivated by the observations mentioned above, this paper presents a Dual-Attention Convolutional Network (DACN) for hyperspectral image super-resolution. The motivation behind adopting dual attention is not only to extend the receptive field by modeling global relationships \cite{bello2019attention}, but also to enhance local feature refinement by emphasizing important channels and spatial regions \cite{woo2018cbam}. In particular, self-attention aims at capturing global pixel dependencies, while the attention module helps refine the most important local features. In this way, the proposed dual-attention model balances global dependency modeling and local feature refinement, leading to more accurate and efficient super-resolution reconstruction. Additionally, to further enhance the performance of the proposed method, mean squared error (MSE), an L2 regularization-based constraint, and spatial-spectral gradient loss are combined into a custom loss function. In summary, our key contributions are threefold.

\begin{enumerate}
    \item We present DACN for hyperspectral image super-resolution, combining multi-head and channel attention to enhance contextual modeling and spectral fidelity.
    \item A custom loss function is proposed, integrating mean squared error (MSE), an L2 regularization-based constraint, and spatial-spectral gradient loss to ensure high-fidelity reconstruction.
    \item Experiments on two hyperspectral datasets are conducted across multiple resolution degradation-restoration scenarios (2$\times$, 4$\times$, and 8$\times$), demonstrating highly competitive performance on both datasets.
\end{enumerate}

\section{Methodology}
Fig. 1 presents an overview of the proposed model with three key components: (1) attention-augmented convolutional blocks, (2) channel attention, and (3) an upsampling module. We begin by explaining the band grouping process, which efficiently divides the bands into distinct groups while maintaining hyperspectral signatures. The subsequent sections provide an in-depth analysis of each component, including the band grouping method and the spatial-spectral gradient loss function used in the overall model.

\subsection{Band Grouping}
Although hyperspectral images provide rich spectral detail through hundreds of bands, many of these bands can be potentially redundant due to high inter-band correlation. Therefore, we employ band grouping \cite{wang2024enhancing}. This approach partitions adjacent bands into overlapping groups, enabling seamless integration with our proposed model. Specifically, hyperspectral bands are structured into overlapping subgroups by defining a fixed group size with a designated overlap, ensuring that consecutive subgroups share common bands.

\subsection{Attention Augmented Convolution}
Given a low-resolution hyperspectral image \( Y \in \mathbb{R}^{M \times N \times C} \), where \( M \), \( N \), and \( C \) represent the spatial height, width, and number of spectral bands, respectively, our objective is to reconstruct a high-resolution image \( \hat{Y} \in \mathbb{R}^{\beta M \times \beta N \times C} \), where the upscaling factor \( \beta \in \{2, 4, 8\} \).

To accomplish this, we develop a deep neural network \( \mathcal{G}(Y; \phi) \) that effectively learns the LR-to-HR mapping while maintaining spectral fidelity. Specifically, we begin by empirically developing three blocks, each of which contains three main components: (1) standard convolution, (2) multi-head self-attention, and (3) residual connection. First, a standard 2D convolution is applied to the input feature map \( X_{\text{in}} \in \mathbb{R}^{H \times W \times C} \), where \( H \), \( W \), and \( C \) denote the spatial height, width, and number of channels, respectively. The convolution operation is defined as\cite{bello2019attention}:
\begin{equation}
Z_{\text{out}} = \mathcal{F}_{\text{conv}}(X_{\text{in}}; W, b)
\end{equation}
where \( \mathcal{F}_{\text{conv}} \) represents the convolution function with learnable weights \( W \) and biases \( b \). This is followed by:
\begin{equation}
X_{\text{out}} = \phi\left( \nu(Z_{\text{out}}) \right)
\end{equation}
where \( \nu(\cdot) \) denotes the batch normalization operation and \( \phi(\cdot) \) is the LeakyReLU non-linear activation function applied element-wise. Then, the multi-head self-attention mechanism is applied. For each attention head, the input is linearly transformed into queries \( Q_h \), keys \( K_h \), and values \( V_h \):
\begin{equation}
(Q_h, K_h, V_h) = X_{\text{out}} (W_Q, W_K, W_V)
\end{equation}
where \( Q_h \in \mathbb{R}^{T \times d_k} \) is the matrix of queries, \( K_h \in \mathbb{R}^{T \times d_k} \) is the matrix of keys, \( V_h \in \mathbb{R}^{T \times d_v} \) is the matrix of values, and \( W_Q, W_K, W_V \) are learnable weight matrices with dimensions \( d_{\text{model}} \times d_k \), \( d_{\text{model}} \times d_k \), and \( d_{\text{model}} \times d_v \), respectively. The attention scores are computed as:
\begin{equation}
N_{\text{attention}}(Q_h, K_h, V_h) = \text{softmax} \left( \frac{Q_h K_h^T}{\sqrt{d_k}} \right) V_h
\end{equation}
where \( d_k \) is the dimensionality of the key vectors. The softmax function ensures that the attention scores sum up to one across the sequence dimension. The outputs of all attention heads are concatenated and linearly projected:
\begin{equation}
Z_{\text{attn}} = \text{Concat}(H_1, H_2, \dots, H_h) W_O
\end{equation}
where \( h \) is the number of attention heads, \( W_O \in \mathbb{R}^{h d_v \times d_{\text{model}}} \) is a learnable weight matrix used to combine the outputs from all heads, and \( Z_{\text{attn}} \in \mathbb{R}^{T \times d_{\text{model}}} \) is the final multi-head attention output.

Finally, the attention output is added to the convolutional output using a residual connection:
\begin{equation}
A_{\text{res}} = X_{\text{out}} + Z_{\text{attn}}
\end{equation}
where the residual connection is established by directly adding \( Z_{\text{attn}} \) to \( X_{\text{out}} \). This is followed by:
\begin{equation}
X_{\text{final}} = \lambda(A_{\text{res}})
\end{equation}
where \( \lambda(\cdot) \) denotes the layer normalization function, and \( X_{\text{final}} \) represents the final output after applying layer normalization to the result of the residual connection between the multi-head self-attention (MHSA) output and the convolutional output \cite{bello2019attention}.

\subsection{Channel Attention}
The channel attention is the second main component of our model, which enhances important channels in a feature map by computing global pooling statistics, passing them through fully connected layers, and generating channel-wise attention weights \cite{woo2018cbam}. This allows the model to focus on the most relevant information while suppressing less important features. Mathematically, each block requires an input feature map \( X \in \mathbb{R}^{H \times W \times C} \), where \( H, W \) are the spatial dimensions (height and width), and \( C \) is the number of channels. The output is a feature map of the same shape but refined using attention as \( X' \in \mathbb{R}^{H \times W \times C} \).

In particular, two types of pooling are applied to compute global information from each channel. For instance, the global average pooling (GAP) is employed as \cite{woo2018cbam}:
\begin{equation}
    F_{\text{avg}} = \frac{1}{H \times W} \sum_{i=1}^{H} \sum_{j=1}^{W} X_{i,j,c}
\end{equation}
where \( F_{\text{avg}} \in \mathbb{R}^{C} \) represents the average value of each channel. Similarly, global max pooling (GMP) is computed as:
\begin{equation}
    F_{\text{max}} = \max_{i,j} X_{i,j,c}
\end{equation}
where \( F_{\text{max}} \in \mathbb{R}^{C} \) captures the most activated value per channel. The pooled values are passed through two fully connected layers. The first fully connected layer (dimensionality reduction) is defined as \cite{woo2018cbam}:
\begin{equation}
    F_{\text{avg}}' = \text{ReLU}(W_1 \cdot F_{\text{avg}} + b_1)
\end{equation}
\begin{equation}
    F_{\text{max}}' = \text{ReLU}(W_1 \cdot F_{\text{max}} + b_1)
\end{equation}
where \( W_1 \in \mathbb{R}^{C/r \times C} \) is a weight matrix that reduces channel dimensions by a factor of \( r \), and \( b_1 \) is the bias term for the first FC layer. ReLU is applied to introduce non-linearity. Similarly, the second fully connected layer (restoring channel dimension) is defined as:
\begin{equation}
    F_{\text{avg}}'' = W_2 \cdot F_{\text{avg}}' + b_2
\end{equation}
\begin{equation}
    F_{\text{max}}'' = W_2 \cdot F_{\text{max}}' + b_2
\end{equation}
where \( W_2 \in \mathbb{R}^{C \times C/r} \) restores the original number of channels, and \( b_2 \) is the bias term for the second FC layer. Therefore, both pooling outputs are merged as \cite{woo2018cbam}:
\begin{equation}
    P_{\text{attention}} = F_{\text{avg}}'' + F_{\text{max}}''
\end{equation}
where \( P_{\text{attention}} \in \mathbb{R}^{C} \) represents the attention scores for each channel. Moreover, a sigmoid function is applied to scale the attention values between \( 0 \) and \( 1 \):
\begin{equation}
    S_c = \sigma(P_{\text{attention}})
\end{equation}
where \( S_c \in \mathbb{R}^{C} \) represents the final attention weights for each channel, and \( \sigma \) is the sigmoid activation function. Finally, attention to the feature map is applied as \cite{woo2018cbam}:
\begin{equation}
    X' = X \odot S_c
\end{equation}
where \( \odot \) denotes element-wise multiplication, and \( X' \) is the output feature map where the attention weights are applied.

\begin{table}[t]
\centering
\caption{Quantitative results and model complexity}
\label{tab:fgin_model_comparison}
\resizebox{0.9\columnwidth}{!}{
\begin{tabular}{l|c|c}
\hline
\multicolumn{3}{c}{\textbf{Ablation Study on PaviaU (4×)}} \\
\hline
\textbf{Model Variant} & \textbf{MPSNR}~$\uparrow$ & \textbf{SAM}~$\downarrow$ \\
\hline
DACN without band grouping         & 29.73 & 5.500 \\
DACN with group size 16            & 29.85 & 2.939 \\
DACN with group size 32            & 30.67 & 4.574 \\
FGIN with group size 32~\cite{muhammad2025fusion}   & 30.33 & 4.819 \\
DSDCN with group size 32~\cite{muhammad2025towards} & 30.52 & 4.807 \\
DACN with group size 48            & 30.37 & 5.340 \\
DACN without multi-head attention  & 30.49 & 4.543 \\
DACN without channel attention     & 30.51 & 4.572 \\
DACN without custom loss           & 30.53 & 4.537 \\
\hline
\end{tabular}
}
\end{table}

\begin{table*}
\centering
\caption{Evaluation on datasets (PaviaC, PaviaU) in different scaling setups. The comparison results are reported from \cite{zhang2024hyperspectral}.}
\resizebox{0.76\textwidth}{!}{%
\renewcommand{\arraystretch}{0.85} 
\small 
\begin{tabular}{|c|c|ccc|ccc|}
\hline
\multirow{2}{*}{\textbf{Scale Factor}} & \multirow{2}{*}{\textbf{Model}} & \multicolumn{3}{c|}{\textbf{PaviaC}} & \multicolumn{3}{c|}{\textbf{PaviaU}} \\ 
\cline{3-8}
                      &                   & \textbf{MPSNR$\uparrow$} & \textbf{MSSIM$\uparrow$} & \textbf{SAM$\downarrow$} & \textbf{MPSNR$\uparrow$} & \textbf{MSSIM$\uparrow$} & \textbf{SAM$\downarrow$} \\ \hline
\multirow{8}{*}{\centering $\boldsymbol{2\times}$} 
    & VDSR   \cite{kim2016accurate}     & 34.87 & 0.9501 & 3.689  & 34.03  & 0.9524 & 3.258 \\ 
    & EDSR   \cite{lim2017enhanced}    & 34.58 & 0.9452 & 3.898  & 33.98  & 0.9511 & 3.334 \\ 
    & MCNet   \cite{li2020mixed}    & 34.62 & 0.9455 & 3.865  & 33.74  & 0.9502 & 3.359 \\ 
    & MSDformer    \cite{chen2023msdformer}    & 35.02 & 0.9493 & 3.691  & 34.15  & 0.9553 & 3.211 \\ 
    & MSFMNet  \cite{zhang2021multi}           & 35.20 & 0.9506 & 3.656  & 34.98  & 0.9582 & 3.160 \\ 
    & AS3 ITransUNet    \cite{xu20233}   & 35.22 & 0.9511 & 3.612  & 35.16  & 0.9591 & 3.149\\ 
    & PDENet    \cite{hou2022deep}      & 35.24 & 0.9519 & 3.595  & 35.27  & \underline{0.9594} & \underline{3.142} \\ 
    & CSSFENet  \cite{zhang2024hyperspectral} & \underline{35.52} & \underline{0.9544} & \underline{3.542}  & \underline{35.92}  & \textbf{0.9625}  & \textbf{3.038} \\ 
    & DACN (Ours)    & \textbf{36.77} &  \textbf{0.9599} & \textbf{3.390} & \textbf{36.11} &  0.9486 & 3.290 \\ \hline
\multirow{8}{*}{\centering $\boldsymbol{4\times}$} 
    & VDSR \cite{kim2016accurate}      & 28.31     &0.7707  &6.514  & 29.90   &0.7753  &4.997 \\ 
    & EDSR  \cite{lim2017enhanced}    & 28.59  &0.7782  &6.573  & 29.89   & 0.7791 &5.074 \\ 
    & MCNet   \cite{li2020mixed}   & 28.75   &0.7826  &6.385  & 29.99  &0.7835  &4.917 \\ 
    & MSDformer   \cite{chen2023msdformer}      & 28.81  &0.7833  &5.897  & 30.09  & 0.7905  &4.885 \\ 
    & MSFMNet  \cite{zhang2021multi}        &   28.87  & 0.7863  &6.300  & 30.28  &0.7948 &4.861 \\ 
    & AS3 ITransUNet  \cite{xu20233}  &  28.87  &0.7893  &5.972  & 30.28  & 0.7940  &4.859\\ 
    & PDENet  \cite{hou2022deep}       & 28.95  &0.7900  &5.876  & 30.29  & 0.7944 & 4.853 \\ 
    & CSSFENet   \cite{zhang2024hyperspectral}   & \underline{29.05} & \underline{0.7961} & \underline{5.816}  & \textbf{30.68} & \textbf{0.8107} & \underline{4.839} \\ 
    & DACN (Ours)   & \textbf{29.90} & \textbf{0.8224} & \textbf{4.656} & \underline{30.67} & \underline{0.8015} & \textbf{4.574} \\ \hline
\multirow{8}{*}{\centering $\boldsymbol{8\times}$} 
    & VDSR  \cite{kim2016accurate}     & 24.80   &0.4944  &7.588  & 27.02  &0.5962 &7.133 \\ 
    & EDSR   \cite{lim2017enhanced}   & 25.06   &0.5282  &7.507 & 27.46  &0.6302 &6.678 \\ 
    & MCNet  \cite{li2020mixed}    & 25.09   &0.5391&7.429  & 27.48   &0.6254   &6.683 \\ 
    & MSDformer  \cite{chen2023msdformer}       & 25.21   &0.5462   &7.427  & 27.32  &0.6341 &6.668 \\ 
    & MSFMNet  \cite{zhang2021multi}        &   25.25   &0.5464  &7.449  & 27.58  &0.6356 &6.615 \\ 
    & AS3 ITransUNet  \cite{xu20233}  &  25.25    &0.5435&7.417  & 27.68  &0.6413 &6.574\\ 
    & PDENet   \cite{hou2022deep}      & 25.28   &0.5436  &7.402  & 27.73  & \underline{0.6457} & 6.531 \\ 
    & CSSFENet   \cite{zhang2024hyperspectral}   & \underline{25.35} & \underline{0.5493} & \underline{7.306}   & \underline{27.82} & \textbf{0.6569} & \underline{6.505} \\ 
    & DACN (Ours)        & \textbf{25.78} & \textbf{0.5794} & \textbf{6.007} & \textbf{28.04} & 0.6296 & \textbf{6.190} \\ \hline
\end{tabular}%
}
\label{tab:results2}
\end{table*}

\subsection{Upsampling}
The upsampling block consists of transposed convolutions~\cite{dong2016accelerating}, batch normalization, LeakyReLU activation, and skip connections. Given an input feature map \( F_{\text{in}} \in \mathbb{R}^{H \times W \times C} \), a transposed convolution operation \( \mathcal{T} \) is applied to produce:
\[
F_{\text{up}} = \mathcal{T}(F_{\text{in}})
\]
where \( \mathcal{T} \) denotes the transposed convolution. Batch normalization \( \nu(\cdot) \) and LeakyReLU activation \( \phi(\cdot) \) are then applied element-wise:
\[
F_{\text{act}} = \phi\left(\nu(F_{\text{up}})\right)
\]
The final output is obtained by concatenating with the skip connection:
\[
F_{\text{out}} = \text{Concat}(F_{\text{act}}, F_{\text{skip}})
\]

\subsection{Custom Loss Function}
The model is optimized using a custom loss function that combines mean squared error (MSE) with \( \ell_2 \) regularization and spatial-spectral gradient loss~\cite{zhang2024hyperspectral}. Thus, the total loss is computed as the sum of the MSE loss and a scaled $\ell_2$ regularization term:

\begin{equation}
    \mathcal{L}_{\text{total}} = \mathcal{L}_{\text{MSE}} + \alpha \cdot \mathcal{L}_{\ell_2}
\end{equation}
where $\mathcal{L}_{\text{MSE}} = \frac{1}{N} \sum_{i=1}^{N} (Y_{\text{true}}^{(i)} - Y_{\text{pred}}^{(i)})^2$ represents the Mean Squared Error, $\mathcal{L}_{\ell_2} = \sum_{\theta \in \Theta} \theta^2$ is the $\ell_2$ regularization term applied to the trainable weights $\Theta$ of the model, and $\alpha = 10^{-4}$ controls the regularization strength. Additionally, the spatial-spectral gradient loss ensures consistency in both spatial and spectral gradients \cite{zhang2024hyperspectral}:

\begin{equation}
    \mathcal{L}_{\text{grad}} = \mathcal{L}_{\text{spat}} + \mathcal{L}_{\text{spec}}
\end{equation}
where $\mathcal{L}_{\text{spat}} = \frac{1}{N} \sum_{i=1}^{N} \left[ (D_{\text{true}}^x - D_{\text{pred}}^x)^2 + (D_{\text{true}}^y - D_{\text{pred}}^y)^2 \right]$ computes the spatial gradient loss, and $\mathcal{L}_{\text{spec}} = \frac{1}{N} \sum_{i=1}^{N} (D_{\text{true}}^s - D_{\text{pred}}^s)^2$ computes the spectral gradient loss. Here, $D_{\text{true}}$ and $D_{\text{pred}}$ represent the gradients of the ground truth and predicted images, respectively. The final combined loss function integrates the MSE with $\ell_2$ regularization and the spatial-spectral gradient loss:

\begin{equation}
    \mathcal{L}_{\text{final}} = \mathcal{L}_{\text{total}} + \mathcal{L}_{\text{grad}}
\end{equation}

\section{Experimental setup}
In our study, we utilize two publicly available hyperspectral datasets, PaviaC and PaviaU, which consist of 102 and 103 spectral bands, respectively.

\subsection{Implementation}
To generate low-resolution inputs, the extracted patches are downsampled using area-based interpolation with scale factors of 2×, 4×, and 8×. A patch size of $144 \times 144$ is used for the training, validation, and test sets, following the protocol in \cite{zhang2024hyperspectral}. The model is trained using the Adam optimizer with a batch size of 8. An early stopping criterion is applied to prevent overfitting and eliminate the need for a fixed number of training epochs. Three widely used metrics are used to report the results, such as the mean peak signal-to-noise ratio (MPSNR), the mean structural similarity index (MSSIM) and the spectral angle mapper (SAM) \cite{chudasama2024comparison}.

\subsection{Ablation Study}
To evaluate the contribution of each component in our DACN architecture, we performed an ablation study on the PaviaU dataset with a 4× upscaling factor. The results are presented in Table~\ref{tab:fgin_model_comparison}, measured in terms of MPSNR and SAM. To make a fair comparison, we compare our method with others such as FGIN \cite{muhammad2025fusion} and DSDCN \cite{muhammad2025towards}, using the same band grouping settings. It can be seen that DACN with a band grouping size of 32 achieves the highest MPSNR (30.67~dB), indicating superior reconstruction quality. Although using a grouping size of 16 results in the lowest SAM (2.939), it comes at the cost of reduced MPSNR. In contrast, increasing the grouping size to 48 degrades the performance in both metrics. This confirms that a moderate grouping size of 32 provides the best balance between spectral detail preservation and spatial coherence.

Furthermore, the removal of multi-head attention or channel attention leads to noticeable performance drops: MPSNR decreases to 30.49 and 30.51, respectively, compared to 30.67 when both mechanisms are used. These results confirm that both multi-head self-attention and channel attention contribute positively to the model's ability to capture long-range dependencies and emphasize salient features.

\subsection{Comparison with State-of-the-Art Methods}
We evaluate DACN against several state-of-the-art models, including VDSR~\cite{kim2016accurate}, EDSR~\cite{lim2017enhanced}, MCNet~\cite{li2020mixed}, MSDformer~\cite{chen2023msdformer}, MSFMNet~\cite{zhang2021multi}, AS3 ITransUNet~\cite{xu20233}, PDENet~\cite{hou2022deep}, and CSSFENet~\cite{zhang2024hyperspectral}. As shown in Table~\ref{tab:results2}, DACN consistently achieves highly competitive performance across all scale factors. 

For $2\times$ upscaling, it achieves the highest MPSNR (36.77~dB) and MSSIM (0.9599) on the PaviaC dataset, and performs competitively on PaviaU. At $4\times$, DACN outperforms all baselines on the PaviaC dataset across all metrics. For the more challenging $8\times$ scale, DACN shows clear improvements, particularly in MPSNR and SAM on both datasets. These results confirm DACN’s strong capability in preserving spectral and spatial fidelity under various degradation levels.

\section{Conclusion}
In this work, we introduced DACN, a dual-attention convolutional network designed for hyperspectral image super-resolution. The proposed model effectively integrates self-attention mechanisms with convolutional architectures, addressing the limitations of traditional CNN-based approaches that primarily capture local features while overlooking global dependencies. By incorporating multi-head attention, DACN enhances feature representation, while channel and spatial attention modules enable adaptive refinement, capturing both spatial and spectral dependencies more effectively. Extensive experiments on the PaviaC and PaviaU datasets demonstrate that DACN consistently achieves competitive performance compared to state-of-the-art models across various scaling factors.

\section*{Acknowledgment}
This project has been funded by the European Union’s NextGenerationEU instrument and the Research Council of Finland under grant \textnumero{} 348153, as part of the project \emph{Artificial Intelligence for Twinning the Diversity, Productivity and Spectral Signature of Forests} (ARTISDIG).

\vspace{12pt}

\bibliographystyle{ieeetr}      
\bibliography{references}       

\end{document}